\documentclass[twocolumn,preprintnumbers,amsmath,amssymb,showpacs]{revtex4}
\usepackage{graphicx,epsfig}   

\begin{document}
\title{
Evolutionary Prisoner's Dilemma in Random Graphs
}

\author
{O. Dur\'an}
\affiliation{``Henri Poincare'' Chair of Complex Systems, Physics Faculty,University of Havana, La Habana, CP10400, Cuba. 
}
\author
{R. Mulet}
\affiliation
{``Henri Poincare'' Chair of Complex Systems and Superconductivity
Laboratory, Physics Faculty-IMRE, University of Havana, La Habana, 
CP 10400,Cuba} 

\date{\today}
\begin{abstract}

We study an evolutionary version of the spatial prisoner's dilemma game, where the agents are placed in a random graph. 
For lattices with fixed connectivity, $\alpha$, we show that for low values of $\alpha$
the final density of cooperating agents depends on the initial conditions, while it does not depend for high connectivity lattices. 
We fully characterized the phase diagram of the system, using both, extensive numerical simulations and analytical computations. 
It is shown that two different behaviors are well defined: 
a Nash equilibrium one, where the density of cooperating agents $\rho_c$ is fixed, and a non-stationary one, where $\rho_c$ fluctuates in time. Moreover we study lattices with fluctuating connectivities and find that the  phase diagram previously developed looses its meaning. In fact, multiple transitions appear and only one regime may be defined. This regime is completely characterized by a non stationary state where the density of cooperating agents varies in time.  
 
\end{abstract}


\maketitle

\section{Introduction}
\label{sec:int}

The search for models able to account for the complex behavior in many biological, economical and social systems has lead to an intense research activity in the last years \cite{pepe}. In particular, a very debated issue is the emergence of cooperation between competitive individuals \cite{Maynard} a problem that was first analyzed by Alxerod \cite{Alxerod} in the context of the Game Theory \cite{book1}.

Game Theory was originally developed to find the optimal strategy for a given game between two intelligent players. However, it´s straightforward development involved the generalization toward the iterated games of $N$ players. In this context many theories have been proposed to explain the emergence and sustainability of cooperation, kin selection \cite{Hamilton}, reciprocal altruism 
\cite{Trivers}, group selection \cite{Silva} and others.

The Prisoner's Dilemma (PD) is the archetype  model of reciprocal altruism \cite{Alxerod_book}. In the game, each player has two options: to defect, or to cooperate. The defector will always have the highest reward $T$ (temptation to defect) when playing against the cooperator which will receive the lowest payoff $S$ (sucker value). If both cooperate they will receive a payoff $R$ (reward for cooperation), and if both defect they will receive a payoff $P$ (punishment).
Moreover, these four payoffs satisfy the following inequalities:

\begin{eqnarray}
\label{PD's-inequalities}
T > R > P > S 
\\
T+S < 2 R \nonumber
\end{eqnarray}

It is not too hard to recognize that for rational players, in a two players-one round game the choice of defection will assure the largest payoff for each player independently of the other decision (Nash Equilibrium) \cite{book2}. This situation, however, creates a dilemma for intelligent players, they know that mutual cooperation results in a higher income for both of them. The question is then under which conditions cooperation emerges in this game.

Nowak and May \cite{Nowak} have shown how cooperation can emerge between players with memoryless strategies in the presence of spatial structure (Spatial Prisoner´s Dilemma, SPD). They considered a deterministic cellular automaton where agents are placed in a square lattice with self, nearest and next-nearest interaction. At each round of the game, the payoff of the player is the sum of the payoffs she got in her encounters with her neighbors. The state of the next generation is defined occupying the site of the lattice with the players having the highest score among the previous owner and the immediate neighbors. It was remarkable the fact that within these simple rules, for a certain range of values of the pay-off matrix, very complex spatial patterns show -up with cooperators and defectors coexisting.

Since then, the game has been largely extended or modified to study more complex situations. For example, Szabo et al. studied the influence of the tic-for-tat strategy 
\cite{Szabo00} and the effects of the external constraint \cite{Szabo02} in the game. Vainstein and Arenzon \cite{Vainstein} approached the problem considering site-diluted lattices to mimic the presence of disorder in the environment and proved that, depending on the amount of disorder, cooperation can be enhanced. Moreover, Abramson and Kuperman \cite{Abramson} and Kim and colaborators \cite{Kim} studied the consequences of different topologies and proved that defectors are enhanced in small-world networks. Furthermore, Ebel and Bornholdt \cite{Ebel} studied the response of the system upon perturbations finding different regimes for avalanche dynamics.

In this work, we try to extend previous results of other authors \cite{long} and to put them in a more general framework. To this end we decided to study the model in a random graph \cite{RGraphs}. The introduction of a random graph may be interpreted as a first step to better characterize the disordered nature of the interactions in evolutionary systems, the extension to more complex networks\cite{RGraphs} will be the subject of a forthcoming work.   
Moreover, while loosing the notion of bi-dimensional or three-dimensional space, we get a very simple way to tune the number of interacting agents.

We study two type of random lattices, in the first one, usually called a Bethe lattice, all the sites have a fixed and equal number of neighbors, $\alpha$ while the other one, is a lattice where the number of links per site is Poissonian distributed with a mean value $\alpha$. 

We show, that for lattices with fixed connectivity the density of cooperating agents develops multiple jumps as a function of the temptation ($T$) of the agents, and that depending of the connectivity these jumps may lead also to a region of the phase space where the density of cooperating agents, $\rho_c$, fluctuate. These transitions are very well characterized and we show that they can be predicted by the simple study of the interaction between clusters of cooperating and defeating agents. However, for Poissonian lattices the situation is more complex, the number of transitions becomes infinite in the thermodynamic limit, and we don't find a stationary state of cooperating agents.

The remaining of the paper is organized as follows. In the next section we present the model with all its details. Then, the numerical results for the
 lattices with fixed connectivities are presented, together with a comparison of the analytical predictions for the phase diagram. In the next section we present the results for lattices with fluctuating connectivities and finally the conclusions are presented.

\section{Model}
\label{sec:mod}

The model is defined by placing two kind of agents, cooperators ($C$) or defectors ($D$), in a random lattice, with fixed or fluctuating connectivities (as mentioned above) and considering that  
the connected pairs interact through the following payoff matrix

\begin{table}[!htb]
\begin{tabular}{l|lr}	
\label{Nowak's-payoff-matrix}
	&$D$	&$C$ \\ \hline
$D$	&0	&$T$ \\ 
$C$	&0	&1 \\ 
\end{tabular}
\caption{Nowak's payoff matrix for one player. $D$-defector, and $C$-cooperator}
\end{table}

\noindent where $C$ stand for cooperator and $D$ for defector and where the temptation $T$ satisfies $1<T<2$ which is consistent with equation (\ref{PD's-inequalities}). 

The agents will interact simultaneously and independently from each other and the agent payoff will be the sum of the payoffs that she wins in her interaction with all her neighbors. 

The evolution of the system proceeds as follows: 
First, each site is occupied by a cooperator ($C$) with a probability $p$,
or by a defector ($D$) with a probability $1-p$. Then, the agents interact following the above payoff matrix, and in the next time step, in every site $i$ of the lattice we will place the agent with the higher payoff between the neighbors of $i$ and $i$ itself. 
The time $t$ is then defined as the number of generations between the current one and the first. The process is repeated, erasing the later payoffs, until the system stabilizes.

In this way our model reproduces a deterministic interaction between memoryless agents. The only source of randomness comes from the lattice structure and the initial conditions $p$, but, as it is shown below, this is enough to produce a very complex behavior. Therefore, the only remaining relevant variables of our problem are the temptation $T$ of the agents and the connectivity $\alpha$ of the lattice. 

Some more variables will be useful in the discussions below, so we will introduce them here. The state ($D$ or $C$) of site $i$ will be characterize by a variable $\theta_i$ that takes the value 1 if the agent is a cooperator and 0 otherwise. In this way, the state of a system with $N$ sites at time $t$ is fully characterized by the set of variables $(\theta_1,\theta_2,\dots,\theta_N)$. We defined also $s_i$ as the number of cooperative neighbors of the agent locate at the site $i$ (by definition $s_i \le \alpha$), and $g_{\theta_i}^{s_i}$ as the total payoff of the agent placed at site $i$ having $s_i$ cooperating neighbors.

A simple analysis of the payoff matrix shows that the agent's payoff will be different from zero only when she plays with at least one 
cooperator. Thus, the agent's payoff depends on the number of her cooperative neighbors. Besides, the agent's payoff also depends on the type of agent, if she is a cooperator, she wins 1, otherwise she wins $T$ for every interaction with a cooperative neighbor. Therefore:

\begin{equation}
\label{eq:local-payoff}
g^{s_i}_{\theta_i} = (T-(T-1)\theta_i)s_i
\end{equation}

Note that for a cooperative agent $g^s_1 = s$, while for a defector one $g^s_0 = T s$ as pointed out before.

Other useful relations follow immediately from equation (\ref{eq:local-payoff}):

\begin{equation}
\label{eq:s-increase}
g^{s}_{\theta} > g^{s-1}_{\theta} \; (1 \le s \le \alpha)
\end{equation}

and,

\begin{eqnarray}
\label{eq:payoff-ext}
g^\alpha_0 & = & \max_{(\theta,s)}g^{s}_{\theta} \nonumber
\\
g^s_0 & > & g^s_1 
\end{eqnarray}

It is important to  think on the system as two populations that invade each other instead of  agents changing their behaviors during the evolution. In fact,
from eq.(\ref{eq:s-increase}), follows that a cooperator surrounded by cooperators has the highest payoff of her neighborhood, therefore from the evolution rules defined above she must continue cooperating. However, if she were free to change her behavior she would defect increasing her payoff and a few time steps later all the system will be defeating and the cooperation will never be an alternative of the players.

Finally before proceed with the presentation of the results, and for future comparisons, let us review the main results obtained when the agents are placed in a square lattice \cite{Nowak,long}.
Using periodic boundary conditions and starting with half of the agents as cooperators and the other half as defectors ($p=0.5$) it was noticed that 
 as function of the temptation, the spatial game has three qualitative different final states: The  first one, for low temptations ($1 < T < 4/3 $), is characterized by a stationary or slightly periodical $\rho_c$ that become a global majority ($\rho_c > 0.5$). The second, for intermediate temptations ($4/3 < T < 3/2$), is characterized by a non stationary $\rho_c(t)$ and spatiotemporal chaos \cite{long,Nowak}, and a third one, for high temptations ($3/2 < T < 2$), characterized again by a stationary or slightly periodical $\rho_c$ but that is now a global minority ($\rho_c < 0.5$).

\section{Fixed connectivity lattices}
\label{sec:fix-con-lat}

\subsection{Phase Diagram}

\begin{figure}[!htb]
\includegraphics[width=0.94 \columnwidth]{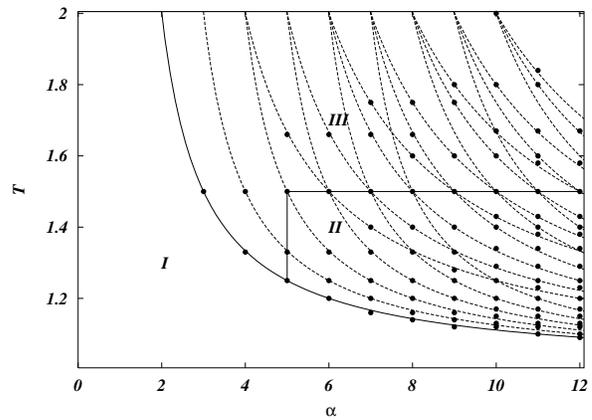}
\caption[0]{ Phase Diagram for the PD game in a fixed connectivity case.
The dashed lines (analytic) and the symbols (simulations) represent the points where $\rho_c$ changes its value. 
The full lines divide three different zones: {\bf I} High values of 
$\rho_c$, {\bf II} Non-stationary values of $\rho_c$ and {\bf III} 
Low values of $\rho_c$. 
}
\label{fig:phase-diagram}
\end{figure}

In figure 
\ref{fig:phase-diagram} it is shown the phase diagram of the model obtained from the simulations (black symbols) and from analytical computations (dashed lines to be discussed below)
when the agents are initially distributed with probability $0<p<1$ 
in the lattice.
The black symbols represent critical temptations ($T_c$) i.e. the temptations values at which $\rho_c$ jumps for each connectivity $\alpha$. 
 Note the perfect coincidence between the points and the predictions, and also the fact that the phase diagram does not depend on $p$, provided of course it is different from 0 and 1.

Three different regimes are very well defined in this figure by the full lines. The first one ({\bf I}) is characterized by a stationary $\rho_c$, the highest for these connectivities $\alpha$ but not necessarily the global majority, as in the two dimensional square lattice. A second ({\bf II}) is characterized by non-stationary states. These states does not necessarily emerge with probability 1. It means, for a given $\alpha$ and a given $T$ in this zone, it will depend on the  initial distribution of the cooperators, and on the particular graph whether this phase is observed or not. And a third ({\bf III}) regime that appears for high values of $T$, is characterized by a stationary $\rho_c$ that is the global minority ($\rho_c<0.5$).

\begin{figure}[!htb]
\includegraphics[width=0.94 \columnwidth]{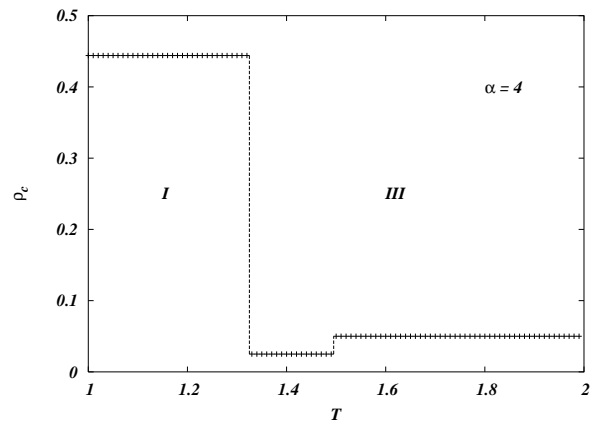}
\caption[0]{$\rho_c$ vs $T$ for random graphs with a fixed connectivity 
$\alpha=4$. {\bf I} and {\bf III} characterize the zones of the phase diagram, (see fig. \ref{fig:phase-diagram}). 
}
\label{fig:rho-vs-Tc4}
\end{figure}

\begin{figure}[!htb]
\includegraphics[width=0.94 \columnwidth]{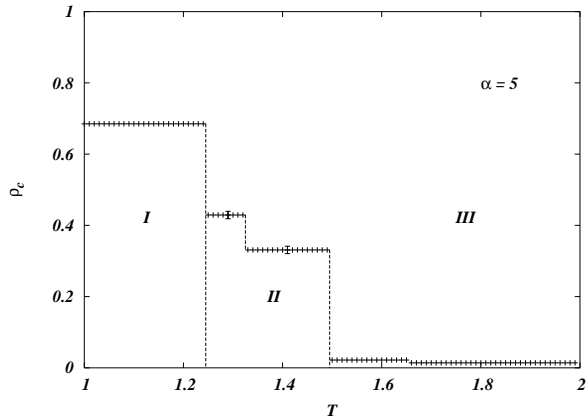}
\caption[0]{$\rho_c$ vs $T$ for random graphs with a fixed connectivity $\alpha=5$.
{\bf I}, {\bf II} and {\bf III} characterize the zone of the phase
 diagram, see (fig. \ref{fig:phase-diagram}). In the zone {\bf II} $\rho_c$ is the mean value of $\rho_c(t)$ and the bars on the lines reflects the fluctuations of $\rho_c$ in this zone.
}
\label{fig:rho-vs-Tc5}

\end{figure}

Note that for a pure PD's game ($\alpha = 1$) and for a one-dimensional chain ($\alpha =2$) there are not transitions. For $\alpha>2$ one can define more than one regime, and within each regime one can observe many transitions, that correspond to jumps in the value of $\rho_c$. To be illustrative about this point, figures \ref{fig:rho-vs-Tc4} and \ref{fig:rho-vs-Tc5} show the variation of $\rho_c$ as a function of the temptation $T$ for connectivities $\alpha=4$ and $\alpha=5$ respectively.

For $\alpha=4$ (fig. \ref{fig:rho-vs-Tc4}) two regimes are present {\bf I} and {\bf III}, and two jumps 
in $\rho_c$. The first jump, indeed, reflects the transition from phase {\bf I} to phase {\bf III} occurring at $T=4/3$ (to be derived below), but the second one, at $T=3/2$, is a jump inside the phase {\bf III}, see figure \ref{fig:phase-diagram}. 
For $\alpha=5$ (fig. \ref{fig:rho-vs-Tc5}) the system is characterized by three different phases and by four jumps in $\rho_c$, two of them are associated to the phase transitions {\bf I-II} at $T=5/4$ and {\bf II-III} at $T=3/2$, while the other two ($T=4/3$ and $T=5/3$) are just jumps of $\rho_c$ within the regimes, see again figure \ref{fig:phase-diagram} to locate these points.

Let us now, try to understand and to deduce the appearance of these jumps and then to explain the classification of the regimes presented above. 

It is not hard to realize that all the dynamics of the model is enclosed in the competition between cooperators and defectors. The cooperators (because of their small payoffs), to survive, must be organized in clusters. So, we may imagine the system as a set of cooperating clusters embedded in a sea of defectors. Then, if the boundary of the cooperator's cluster is strong enough it will grow, otherwise, the cluster keeps its size or becomes smaller.

Following this analysis, it is important to point out that in this model only phases with cooperators and defectors coexisting may appear. In fact, the defective population can invade the whole graph ($\rho_c=0$) if the lowest possible payoff for a boundary defector $g_0^1$, i.e a defector interacting with only one cooperator, is larger than the highest possible payoff of a cooperator 
 $g_1^\alpha$. From equation (\ref{eq:local-payoff}) this may happen if $T>\alpha$, but by definition $T<2$, therefore $\rho_c > 0$ for all $\alpha \ge 2$. For $\alpha=1$ the result is obviously the original Prisoner Dilemma. 

Moreover, a whole invasion of the graph by cooperating agents is impossible because one defector surrounded by cooperators will have the highest possible payoff $g_0^{\alpha}$, see equation (\ref{eq:payoff-ext}), and is therefore indestructible, $\rho_c<1$.

Once the connectivity of the graph is fixed, it is interesting to see what happens when the temptation $T$ of the agents increases. Obviously, if $T=1$ the systems does not evolve in time, it keeps its initial distribution of cooperators and defectors.
Once the temptation increases, we may ask ourselves, at which value of $T$ will this distribution change, or more generally at which value of $T$ $\rho_c$ is going to change . In order to find such a behavior it is evident that either the cooperating clusters most become soft at their boundaries such that they get invaded by the defectors, or they most become strong enough to occupy part of the sea of defectors. Then, the condition of equilibrium that most be satisfied by all the agents in the boundary between cooperators clusters and defectors is the following:

\begin{equation}
\label{eq:equil}
g^{s_1}_1 = g^{s_0}_0  
\end{equation}

\noindent where $s_1$ and $s_0$ stands for the number of cooperative neighbors of the cooperator and the defector respectively. But from equation (\ref{eq:local-payoff}) $g_1^{s_1}=s_1$ and $g_0^{s_0}= T s_0$. Then,

\begin{equation}
\label{eq:Tc}
T_c = \frac{s_1}{s_0} \nonumber
\end{equation}

\noindent and since we are interested in the region $T>1$, the equation (\ref{eq:Tc}) implies that $s_1>s_0$. Therefore we may substitute $s_1=\alpha-n$ and 
$s_0=\alpha-n-m$ in equation (\ref{eq:Tc}) to get:

\begin{equation}
\label{eq:Tc-nm}
T_{c_{n,m}}(\alpha) = \frac{\alpha-n}{(\alpha-n)-m}  
\end{equation}

\noindent where obviously $n<\alpha$ and since $T<2$, $m$ must also satisfy:

\begin{equation}
\label{eq:m-def}
1 \le m \le \mbox{int} \left( \frac{(\alpha-n)-1}{2}\right)
\end{equation}

In this way, assigning appropriate values of $n$ and $m$ to the equation (\ref{eq:Tc-nm})  we characterize all the jumps of $\rho_c$ for a given $\alpha$ as a function of $T$ . This is what is represented in figure \ref{fig:phase-diagram} by dashed lines. Of course, the lines are just guided to the eyes, and fixed values of $\alpha$ should be understood when analyzing equation (\ref{eq:Tc-nm}). 

Going deeper in this kind of analysis we may see that, from equations (\ref{eq:s-increase}) and (\ref{eq:payoff-ext}), the strongest cooperators hold:

\begin{equation}
\label{eq:phase-1}
g^s_1 > g^{s-1}_0 \; (2 \le s \le \alpha)
\end{equation}

\noindent which means that the defectors can not invade the clusters of cooperators, or viceversa a cooperator surrounded by $s$ cooperators will invade all the defectors neighbors with less than $s$ cooperators around them. In other words, while equation (\ref{eq:phase-1}) holds the clusters of cooperators grow inside the sea of defectors (except for $\alpha = 2$, when the winner cooperator does not have any defector to invade). At some point the defectors loneliness increases enough and they become indestructible, stopping the propagation of cooperators.

Then, following (\ref{eq:local-payoff}), the set of inequalities (\ref{eq:phase-1}) are satisfied for the temptations:

\begin{equation}
\label{eq:T-1}
T < \frac{\alpha}{\alpha-1}
\end{equation}

\noindent that correspond to the first transition at the lowest $T_c(\alpha)$, from equation (\ref{eq:Tc-nm}). 
Below this line, appears what we call regime {\bf I}, there, $\rho_c$ evolves toward a stationary state where it reaches, depending on the initial conditions,  its highest possible value.

On the other hand, for the temptation range:

\begin{equation}
\label{eq:Tc-3}
\frac{3}{2} < T < 2 
\end{equation}

\noindent the opposite condition is satisfied, i.e:

\begin{equation}
\label{eq:relax-2}
g^1_0 < g^s_1 < g^{s-1}_0 \; (3 \le s < \alpha)
\end{equation}

\noindent independently on the initial conditions and the connectivity of the lattice. Now, the spreading of cooperators over defective sites is strongly reduced and defectors dominate the system. The condition (\ref{eq:relax-2}) implies again a stationary final state. But, due to the greatest domination of defectors, $\rho_c$ reaches its  lowest possible value for the given initial conditions (regime {\bf III}).

In the intermediate range appears what we call regime {\bf II}. In this regime 
the stability conditions (\ref{eq:phase-1}) and (\ref{eq:relax-2}) with an absolute winner do not hold anymore and dynamics instabilities appear in the interior of the lattice. Depending on the temptation, $T$, boundary sites are intermittently occupied by defectors or cooperators, or alternatively lines of defectors travel across the cluster of cooperators.

Moreover, note that in the phase diagram the regime {\bf III} is not only limited to values of $T$ larger than $3/2$. For $\alpha=4$ the regime {\bf II} is not present, a surprising result if one consider that it is present in the square lattice. Unfortunately we were not able to analytically justify this fact, but, we are tempted to conjecture that this is a consequence of the absence of spatial correlations in the Bethe lattice.

Summarizing, from equations (\ref{eq:T-1}) and (\ref{eq:Tc-3}) 
we may divide the phase diagram in three zones, represented by full lines in figure \ref{fig:phase-diagram}:

\begin{eqnarray}
\label{regimes}
{\bf I :}& 1 < T < \frac{\alpha}{\alpha-1} \nonumber
\\
{\bf II :}& \frac{\alpha}{\alpha-1} < T < \frac{3}{2} \hspace{0.2in} (\alpha>4)
\\
{\bf III :}& \frac{3}{2} < T < 2 \nonumber \hspace{0.2in} \mbox{if $\alpha \ne 4$}
\\
{\bf III :}& \frac{\alpha}{\alpha-1} < T < 2  \hspace{0.2in} \mbox{if $\alpha=4$} \nonumber
\end{eqnarray}

Another interesting question to be answer is how many transitions we will find for a fixed connectivity when the temptation changes.

It could be seen from figure \ref{fig:phase-diagram}, that at a given connectivity $\alpha_o$ you will find all the transitions observed in all the previous $\alpha$'s, i.e., for  $\alpha_o>\alpha$ (equation (\ref{eq:Tc-nm})).
In order to calculate the number of transitions to each connectivity ($f(\alpha)$), note that,  neglecting repetitions, there are only $int(\alpha/2)$ new $T_c$ values  when one moves from 
$\alpha$ to $\alpha + 1$, equation (\ref{eq:m-def}). But:

\[\mbox{int}\left( \frac{\alpha}{2}\right) = \left\{ \begin{array}{ll} 
						\frac{\alpha}{2} & \mbox{$\alpha$ even} \\
						\frac{\alpha-1}{2} & \mbox{$\alpha$ odd}
						\end{array}
					\right. \]

\noindent Then, taking the average: 

\[
\left< \mbox{int}\left( \frac{\alpha}{2}\right) \right> = \frac{\alpha-(1/2)}{2} 
\]

\noindent we get the incremental equation

\begin{equation}
f(\alpha + 1) - f(\alpha) = \frac{\alpha-(1/2)}{2}  
\label{eq:incr-eq}
\end{equation}

For $\alpha > 1$ and taking the initial condition $f(1) = 0$, the solution of (\ref{eq:incr-eq}) is:

\begin{equation}
\label{eq:f1}
f(\alpha) =  \frac{(\alpha-1)^2}{4}   
\end{equation}

\begin{figure}[!htb]
\includegraphics[width=0.94 \columnwidth]{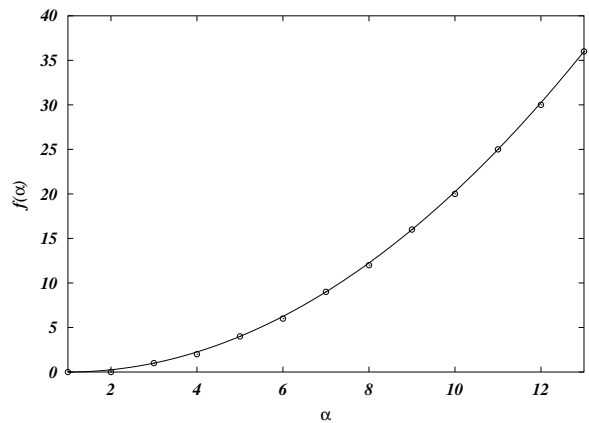}
\caption[0]{The number of $\rho_c$ transitions for a lattice with a fixed connectivity $\alpha$. The points represent the simulation results and the line the analytical function $f(\alpha)$ from equation (\ref{eq:f1}).
}
\label{fig:num-states}
\end{figure}

A quadratic expression that compares perfectly with the results of the simulations (see fig.\ref{fig:num-states}).

\subsection{Connectivity and initial conditions dependence}

In the previous subsection we presented numerical simulation and arguments that justify the independence of the phase diagram from the initial conditions. Here, we go a step further trying to understand how $\rho_c$ changes with the initial conditions and the connectivity in the different regimes of the phase diagram. 

\begin{figure}[!htb]
\includegraphics[width=0.94 \columnwidth]{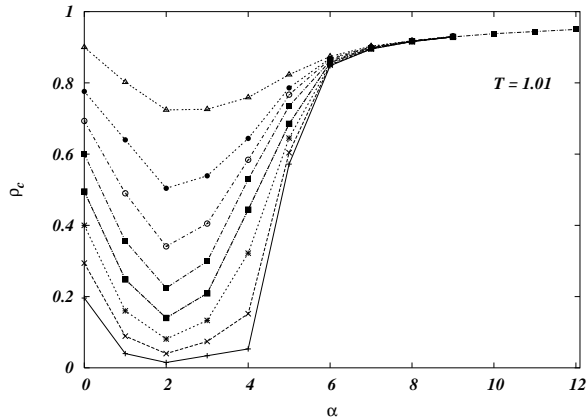}
\caption[0]{Initial conditions dependence of $\rho_c$ in regime {\bf I}. 
From bottom to top  $p$ varies from $0.2$ to $0.9$ in steps of $0.1$. Note the convergence of $\rho_c$ when $\alpha > 5$ and the minimum reached by the one dimensional chain.
}
\label{fig:init-cond-I}
\end{figure}

\begin{figure}[!htb]
\includegraphics[width=0.94 \columnwidth]{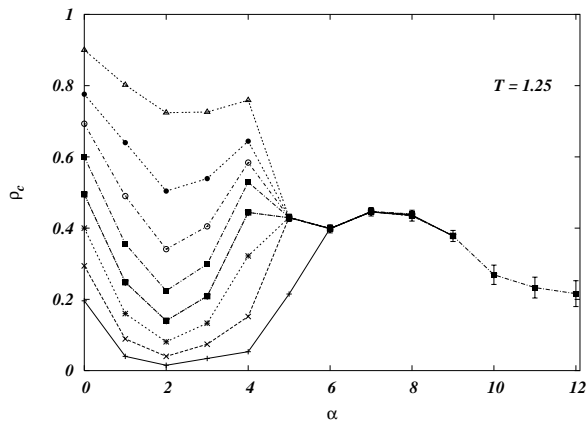}
\caption[0]{Initial conditions dependence of $\rho_c$ at $T = 5/4$, regime {\bf II}. From bottom to top  $p$ varies from $0.2$ to $0.9$ in steps of $0.1$. Note the convergence of $\rho_c$ when $\alpha \ge 5$.
}
\label{fig:init-cond-II}

\end{figure}

\begin{figure}[!htb]
\includegraphics[width=0.94 \columnwidth]{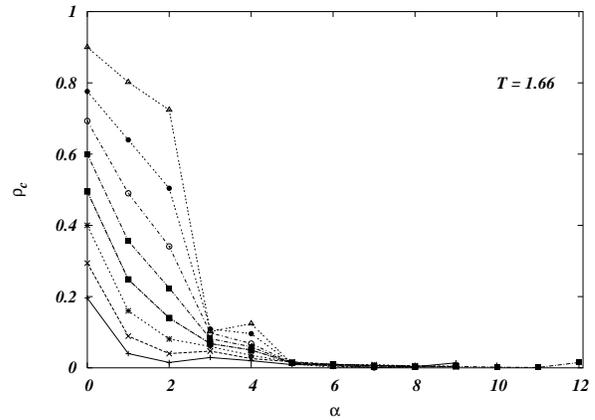}
\caption[0]{Initial conditions dependence of $\rho_c$ at $T = 5/3$, a characteristic $T_c$ in the third phase.  Note again the convergence of $\rho_c$ when $\alpha \ge 5$.
}
\label{fig:init-cond-III}

\end{figure}

The main interesting result to be shown is the dependence of the density of cooperating agents with the initial conditions as a function of the connectivity.
This is shown in figures  \ref{fig:init-cond-I}, \ref{fig:init-cond-II} and \ref{fig:init-cond-III} where each figure represents results for values of $T$ characteristics of the different regimes of the system. In each figure, the curves show for different values of $p$ the final value of $\rho_c$. In figure \ref{fig:init-cond-I} and \ref{fig:init-cond-III} $\rho_c$ is a stationary value. In figure \ref{fig:init-cond-II}, $\rho_c$ is the average value of the density of cooperating agents in a fluctuating state (regime {\bf II}), and the bars reflect the fluctuations of $\rho_c$.

As all the figures show, independently of $p$, the curves follow similar patterns for the corresponding values of $T$.

In figure \ref{fig:init-cond-I}, we find, first, a sharp decrease of $\rho_c$ that reaches its minimum for graphs with connectivity $2$ (that represent the one-dimensional chain) and then $\rho_c$ increases quickly up to $\alpha=6$ and then slower for higher connectivities. Remarkable the fact, that all the curves are identical for values of $\alpha$ greater than $5$. 

In figure \ref{fig:init-cond-II} two different behaviors are very clear, first for low connectivities we obtain curves similar to those of figure \ref{fig:init-cond-I} a result easily explained, in fact for $\alpha<5$ the system is still in regime {\bf I}. It is more relevant the fact that for $\alpha \ge 5$ all the results         coincide (disregarding $p=0.2$, that is probably  strongly affected by finite size effects). Moreover the mean value of $\rho_c$ decreases for higher connectivities while its fluctuations increase.

Finally, figure \ref{fig:init-cond-III} represents the behavior of the system in the regime {\bf III}. There, increasing the connectivity $\rho_c$ decreases, and again for $\alpha \ge 5$ it is difficult to distinguish the curves.

From these figures we may extract important conclusions. Independently of $p$ the curves always follow similar patterns that are only defined by the temptation of the agents to defeat, and that $p$ only defines the value of $\rho_c$. Moreover, if the connectivity is high enough, $\alpha>5$, $\rho_c$ becomes also independent of the random initial conditions and the game is fully characterized by the connectivity $\alpha$ and $T$.

To understand the effects of connectivity on the game dynamics, let us define $P^s_\theta (\alpha)$ as the probability of an agent doing $\theta$
to has $s$ cooperative neighbors in a lattice with connectivity $\alpha$. As before $\theta=1$ reflects cooperation and $\theta=0$ reflects defection.
 
Based on combinatorial arguments we find that:

\begin{equation}
\label{eq:P-definition}
P^s_\theta (\alpha) = \left( \begin{array}{c} \alpha \\ s \end{array} \right) p^s (1-p)^{\alpha-s} (p \theta + (1-p)(1-\theta))
\end{equation}

Then, the following relations hold

\begin{equation}
\label{eq:relation-1}
P^s_\theta (\alpha) > P^s_\theta (\alpha-1) \hspace{0.1in} \mbox{if $s > \alpha p$}
\end{equation}

\begin{equation}
\label{eq:relation-2}
\left. 	
	\begin{array}{l}	
	P^{s+1}_\theta (\alpha) > P^s_\theta (\alpha)
	\\
	P^{s+1}_\theta (\alpha) > P^s_\theta (\alpha-1) 
	\end{array}
\right \}
	\hspace{0.1in}	\mbox{if $s < \alpha p-1$} 
\end{equation}

Both relations tell us that, independently of $p$, the probability to have $s$ or more cooperative neighbors increases when the connectivity grows.
Thus, for a given initial condition, a larger connectivity implies an increase of the number of cooperators linked to both kind of agents. This leads to a local payoff increment that amplify the domination of the winners agents in each phase, therefore, explaining the behaviors shown in figures \ref{fig:init-cond-I}, \ref{fig:init-cond-II} and \ref{fig:init-cond-III}.

In regime {\bf I}, as the cooperator's clusters can only grow, they are strongly enhanced when the connectivity increase. This is clearly what is shown in 
figure  \ref{fig:init-cond-I}, where for $\alpha>2$ $\rho_c$ continuously increases. 
On the contrary, in the regime {\bf III} cooperators can hardly invade the defectors, who win practically all boundary interactions (see equation (\ref{eq:relax-2})). 
Then, from relations (\ref{eq:relation-1}) and (\ref{eq:relation-2}), a connectivity increment virtually exterminates all the cooperators in the system (see figure \ref{fig:init-cond-III}). Furthermore in regime {\bf II}, both kind of agents are enhanced when the connectivity increases.
This has two effects: the first is to increase the cooperator's density fluctuations  and the second to slightly reduce the mean value of the density of cooperators (figure \ref{fig:init-cond-II}). Note also, that in figure \ref{fig:init-cond-II}, for $\alpha<5$ we are in the regime {\bf I} and therefore the explanation above applies.

For $\alpha=1$ we have the simple Prisoner's Dilemma Game, and for $\alpha=2$ the analytic solution of the $\rho_c$ versus $p$ is calculated in the Appendix of this work.

\section{Fluctuating connectivity lattices}
\label{sec:fluc-con-lat}

To study lattices with fluctuating connectivities, we assign to each vertex of the lattice a number of links determined by a Poissonian distribution with mean $\alpha$:

\begin{equation}
\label{eq:poisson}
P(\alpha_i) = \frac{\alpha^{\alpha_i}}{\alpha_i{\bf !}} \exp{(-\alpha)}
\end{equation}

For this lattices the local equilibrium conditions satisfied by neighboring and opposite agents remains:

\[g^{s_1}_1 = g^{s_0}_0\]

\noindent with the only difference that now the connectivities $\alpha_0$ and $\alpha_1$ of both sites may be different, a situation that must be taken in consideration during the analysis of the phase diagram.

Following the analysis done for the fixed connectivity case
it is easy to realize that the following temptations characterize the equilibrium conditions for the boundary war between cooperating and defeating agents. 

\begin{equation}
\label{eq:local-Tc}
T_{c_{n,m}}(\alpha_i) = \frac{\alpha_i-n}{(\alpha_i-n)-m}  
\end{equation}

\noindent where $\alpha_i = \max \left\{ \alpha_0,\alpha_1 \right\}$,
and $T_{c_{n,m}}(\alpha_i)$ is the critical temptation for all sites with connectivity $\alpha_i$. 

\begin{figure}[!htb]
\includegraphics[width=0.94 \columnwidth]{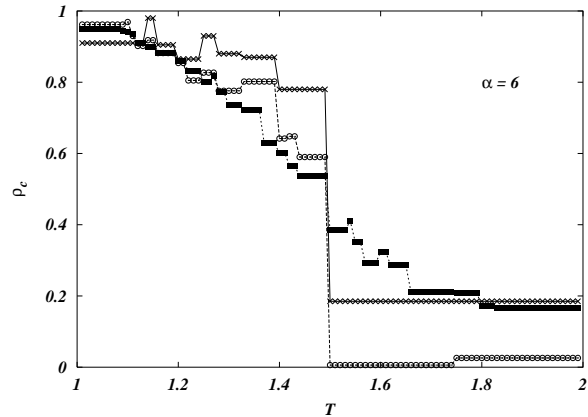}
\caption[0]{$\rho_c$ vs $T$ for different graph's sizes: $N = 100$ (crosses), 1000 (white circles) and 10 000 (black symbols).
}
\label{fig:N-dep}

\end{figure}

The main difference here, comes from the large number of possible connectivities that can be found in this kind of lattices. In fact, equation (\ref{eq:local-Tc}) is more general than (\ref{eq:Tc}). Moreover, the larger the lattice size, the larger the values of $\alpha_i$'s that may be found in the lattice.
Therefore , the number of transitions defined by (\ref{eq:local-Tc}) increases with the lattice size. In the thermodynamic limit $N\rightarrow \infty$, an infinite number of transitions must be expected.
This is shown in figure \ref{fig:N-dep}, where we plot $\rho_c$ vs $T$ for different values of $N$ and $\alpha=6$.

\begin{figure}[!htb]
\includegraphics[width=0.94 \columnwidth]{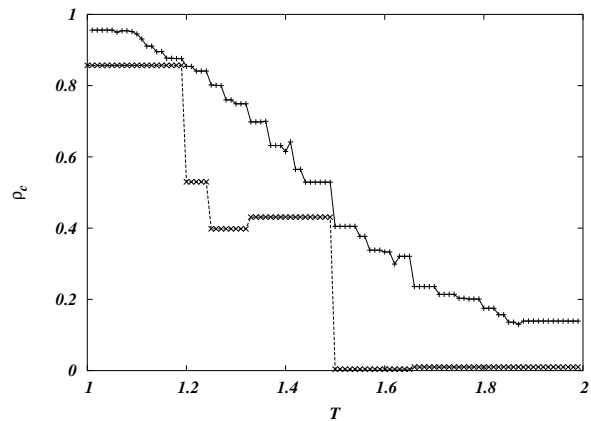}
\caption[0]{Comparison between $\rho_c$ vs $T$ for fixed and fluctuating connectivity $\alpha = 6$. Note that for the latest $\rho_c$ is always greater.
}
\label{fig:fluct-fix}

\end{figure}

Another remarkable result in this kind of lattices is that $\rho_c$ is strongly enhanced with respect to the finite connectivity lattices (see figure \ref{fig:fluct-fix}). This in good agreement with the results of \cite{Vainstein} for square lattices with quenched disorder. However, here it does not reflect the topological accidents of the lattice, but its  random structure. In this kind of lattices it is possible the  
existence of cooperator sites with high connectivities that become the core of a cooperator resistance at high values of $T$.

With respect to the phase diagram, it is important to note that there is not a condition equivalent to the equation  (\ref{eq:phase-1}). In fact there is always a critical temptation lower than $T_c(\alpha_i)$ for any $\alpha > 2$ and therefore the regime {\bf I} must be absent in this lattice. This means that cooperators will never be absolute winners. 
Moreover, cooperative agents placed at sites with the largest connectivities in their neighborhood may resist any growth of the temptation and become seeds for the spreading of cooperation. Therefore also the regime {\bf III} is absent, there is not a stationary phase where defeating agents absolutely dominate over cooperating ones. In short, only a non stationary phase similar to regime {\bf II} is presented in the whole range of $T$.

\begin{figure}[!htb]
\includegraphics[width=0.94 \columnwidth]{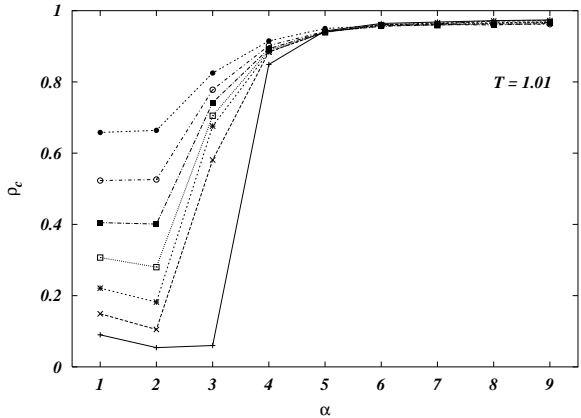}
\caption[0]{Initial condition dependence for fluctuating connectivity lattice at $T = 1.01$. Here $\alpha$ is the mean connectivity and $p$ change from $0.2$ to $0.8$. }
\label{fig:init-cond-cfI}

\end{figure}

\begin{figure}[!htb]
\includegraphics[width=0.94 \columnwidth]{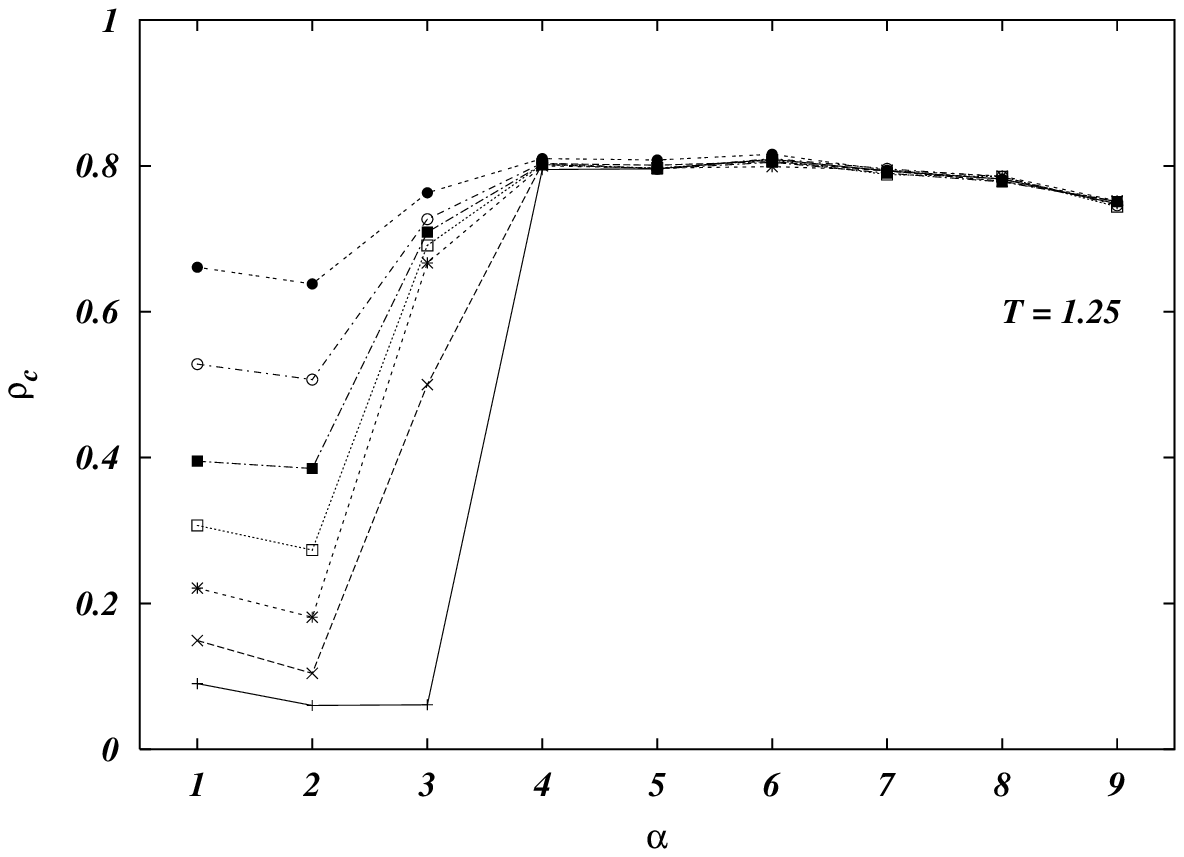}
\caption[0]{Initial condition dependence for fluctuating connectivity lattice at $T = 5/4$. $p$ change from $0.2$ to $0.8$.}
\label{fig:init-cond-cfII}

\end{figure}

\begin{figure}[!htb]
\includegraphics[width=0.94 \columnwidth]{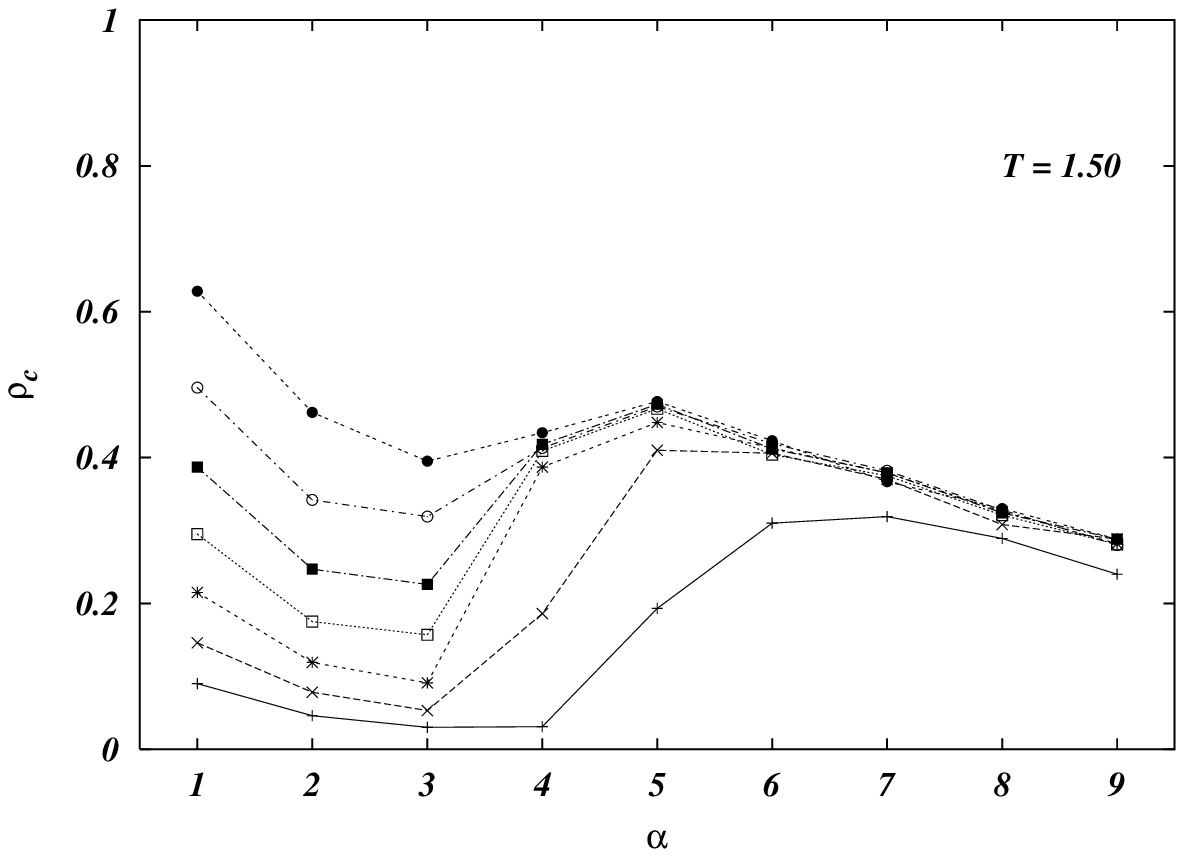}
\caption[0]{Initial condition dependence for fluctuating connectivity lattice at $T = 3/2$. $p$ change from $0.2$ to $0.8$. }
\label{fig:init-cond-cfIII}

\end{figure}

\begin{figure}[!htb]
\includegraphics[width=0.94 \columnwidth]{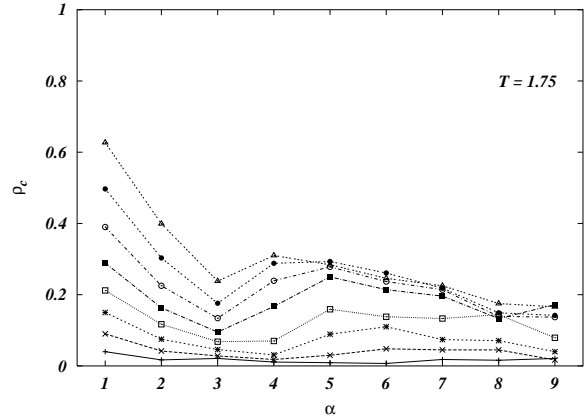}
\caption[0]{Initial condition dependence for fluctuating connectivity lattice at $T = 5/3$. }
\label{fig:init-cond-cfIV}

\end{figure}

Finally, figures \ref{fig:init-cond-cfI}, \ref{fig:init-cond-cfII}, \ref{fig:init-cond-cfIII} and \ref{fig:init-cond-cfIV} show the initial condition and connectivity dependence of $\rho_c$ for different values of $T$.

It is interesting that, despite of the absence of regime {\bf I}, the curves for low temptations are very similar to those for the fixed connectivity lattices. $\rho_c$ becomes independent of $p$ for $\alpha \ge 5$ for small values of $T$.

For high temptations, figure \ref{fig:init-cond-cfIV} shows that $\rho_c$ is highly sensitive to the initial concentration of cooperators. 

The explanation of the connectivity dependence of $\rho_c$ is similar to that for the fixed connectivity lattice, see equations (\ref{eq:relation-1})
and (\ref{eq:relation-2}). The increase of the connectivity enhances the winner agents, therefore, for low temptations, the cooperating agents increase, while for higher values of $T$ the defectors dominate the game and prevent the spreading of cooperators.

\section{Conclusions}
\label{sec:conc}

We present a study of the characteristics of the Spatial Prisoner's Dilemma in random lattices. For lattices with fixed connectivities we were able to fully characterized the phase diagram probing the existence of three different regimes depending on the temptation of the agents and the connectivity of the lattice but independently on the initial conditions of the system. We also give analytical arguments to explain the appearance of these regimes. Furthermore, for these kind of lattices we show that for connectivities larger than $\alpha=5$, also the density of cooperating agents, $\rho_c$, is independent of the initial conditions. Moreover, we give arguments that demonstrate that (in the thermodynamic limit) for lattices with fluctuating connectivities the density of cooperating agents changes continuously with the temptations of the agents. Furthermore, we also showed that in this kind of lattices only the non-stationary regime exists, independently of the temptation, the connectivity of the lattice and the  initial conditions. We have shown that the cooperation is strongly enhanced in comparison with the fixed connectivity lattice.

\appendix*

\section{Estimating the density of cooperating agents}
\label{sec:dens_coop}

It appears clear from equation (\ref{eq:Tc}) and figure \ref{fig:phase-diagram} that there are only 
two connectivities without transitions. For $\alpha=1$ the problem is reduced to the standard Prisoner's Dilemma game, for $\alpha=2$ the agents are essentially located on straight lines.
To find the $\rho_c$ dependence with the initial condition reduces to the calculation of the fraction of initial cooperating agents $p$ that survive in their war with the neighboring defectors, and also to determine the new cooperators that emerge from these interactions.

It is clear that for $\alpha = 1$ all the sites are links by pairs and therefore the interaction reduces to a simple two-person prisoner's dilemma game. 
Since in this game cooperators survive only if they are linked with cooperators, $\rho_c$ will be the probability to find two cooperators linked each other. That is:

\begin{equation}
\label{eq:c1}
\rho_c = p^2
\end{equation}

Note that it is true only if agents, as pointed before, can not change their behavior (i.e. either to cooperate or to defect), in fact, in the original
prisoner's dilemma game the Nash equilibrium is to defect always.

Let us define, $P^{s_1 s_2 \dots s_n}_{\theta_1 \theta_2 \dots \theta_n}$ as the initial probability to have a site $\theta_1$, with $s_1$ cooperative neighbor, linked to a site  $\theta_2$ which  $s_2$ cooperative neighbors linked to a site $\theta_3$ with $s_3$ cooperative neighbors,  and so on, until the site $\theta_n$ with $s_n$ cooperative neighbors.

Using the equation (\ref{eq:P-definition}) we found that for two sites ($n = 2$ in the above definition) this probability is

\begin{eqnarray}
\label{eq:P-2}
P^{s_1 s_2}_{\theta_1 \theta_2}=\left( \begin{array}{c} \alpha \\ s_1 \end{array} \right) \left( \begin{array}{c} \alpha-1 \\ s_2-\theta_1 \end{array} \right) \left( \frac{p}{1-p} \right)^{s_1+s_2-\theta_1} 
\\(1-p)^{2\alpha-1} (p\theta_1 + (1-p)(1-\theta_1))(s_1 \theta_2 + (\alpha-s_1)(1-\theta_2)) 	\nonumber
\end{eqnarray} 

For three sites it could be compute recursively from (\ref{eq:P-2})

\begin{equation}
\label{eq:P-3}
P^{s_1 s_2 s_3}_{\theta_1 \theta_2 \theta_3} = \frac{P^{s_1 s_2}_{\theta_1 \theta_2} P^{s_2 s_3}_{\theta_2 \theta_3}}{P^{s_2}_{\theta_2}} f(\theta_1,s_2,\theta_3)
\end{equation}

\noindent where

\begin{equation}
\label{eq:f}
f(\theta_1,s_2,\theta_3) = \left( \frac{(\alpha-s_2) + (\theta_1-1)}{\alpha-s_2} \right) (1-\theta_3) + \left( \frac{s_2-\theta_1}{s_2} \right) \theta_3
\end{equation}

Similarly an extension of (\ref{eq:P-3}) for general $\theta$'s and $n$'s is:

\begin{equation}
\label{eq:P-n}
P^{s_1 \dots s_n}_{\theta_1 \dots \theta_n} = \frac{P^{s_1 \dots s_{n-1}}_{\theta_1 \dots \theta_{n-1}} P^{s_{n-1} s_n}_{\theta_{n-1} \theta_n}}{P^{s_{n-1}}_{\theta_{n-1}}} f(\theta_{n-2},s_{n-1},\theta_n)
\end{equation}

For $\alpha = 2$ we calculate $\rho_c(p)$ using a general recursive framework which proceed as follows: first, at each generation $t$ we found the density of cooperators with $s$ ($0 \le s \le 2$) cooperative neighbors  taking into account the density of cooperators in the preceding generation and the number of those sites created or destroyed in the current one. 
After a very cumbersome but straightforward algebra we obtain:

\begin{equation}
\label{eq:s}
\rho_{c} = P^{2}_{1} + P^{12}_{11} - P^{212}_{110} - \frac{1}{2} P^{21212}_{01110} - 2 P^{11212}_{01110} - P^{212212}_{011110}
\end{equation}

\noindent and using the equations (\ref{eq:P-definition}), (\ref{eq:P-2}) and (\ref{eq:P-n}) we may compute these probabilities obtaining $P^2_1 = p^3$, $P^{12}_{11} = 2p^3(1-p)$, $P^{212}_{110} = 2p^4(1-p)$, $P^{21212}_{01110} = 2p^5(1-p)^2$, $P^{11212}_{01110} = 2p^4(1-p)^3$ and $P^{212212}_{011110} = 2p^6(1-p)^2$.

\noindent Then $\rho_c$ as function of $p$ becomes:

\begin{equation}
\label{eq:c2}
\rho_c(p) = p^3 (3 - 8p + 13p^2 - 12p^3 + 7p^4 - 2p^5) 
\end{equation}

\begin{figure}[!htb]
\includegraphics[width=0.94 \columnwidth]{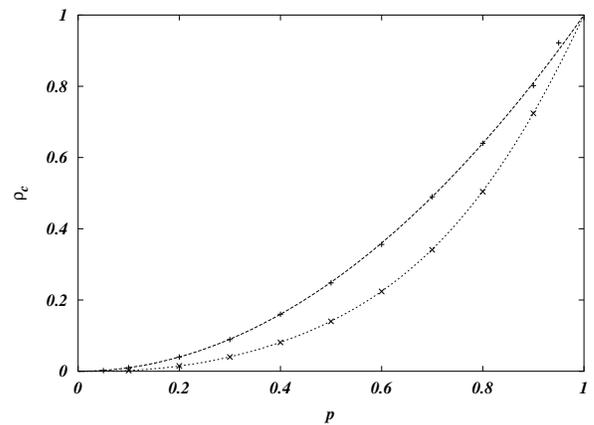}
\caption[0]{Analytical fit of simulation results for $\rho_c$ as function of $p$ for $\alpha =1$ ($+$) and $\alpha =2$ ($x$)}
\label{fig:coop-1-2}

\end{figure}

Both equations, (\ref{eq:c1}) and (\ref{eq:c2}), are represented in figure \ref{fig:coop-1-2}, fitting the simulation results.

\end{document}